\documentclass[aps,pra,reprint,showpacs,amsmath,amssymb,10pt]{revtex4-1}
\usepackage{graphicx}% Include figure files
\usepackage{bm}% bold math
\usepackage{slashed}

\begin{document}
\title{Supercontinuum and ultra-short pulse generation from nonlinear Thomson and Compton scattering}
\author{K. Krajewska$^\dagger$}
\email[E-mail address:\;]{Katarzyna.Krajewska@fuw.edu.pl}
\author{M. Twardy$^\ddagger$}
\author{J. Z. Kami\'nski$^\dagger$}
\affiliation{$^\dagger$Institute of Theoretical Physics, Faculty of Physics, University of Warsaw, Ho\.{z}a 69,
00-681 Warszawa, Poland \\
$^\ddagger$Faculty of Electrical Engineering, Warsaw University of Technology,
Pl. Politechniki 1, 00-661 Warszawa, Poland}

\date{\today}

\begin{abstract}
Nonlinear Thomson and Compton processes, in which energetic electrons collide with an intense optical pulse, are investigated in the framework of classical and
quantum electrodynamics. Spectral modulations of the emitted radiation, appearing as either oscillatory 
or pulsating structures, are observed and explained. It is shown that both processes generate a bandwidth radiation 
spanning the range of few MeV, which occurs in a small cone along the propagation direction of the colliding electrons. Most 
importantly, these broad bandwidth structures are temporarily coherent which proves that Thomson and Compton processes
lead to generation of a supercontinuum. It is demonstrated that the radiation from the supercontinuum can be
synthesized into zeptosecond (possibly even yoctosecond) pulses. Thus, confirming that Thomson and Compton scattering can be used as novel 
sources of an ultra-short radiation, opening routes to new physical domains for strong laser physics. 
\end{abstract}

\maketitle

\section{Introduction}

Conventionally, a high-energy radiation has been produced in large-scale electron accelerators, which has resulted in a pulsed synchrotron 
radiation lasting for picoseconds. More compact sources of high-energy radiation have been built based on
laser-wakefield electron acceleration~\cite{Tajima}. The latter typically produce femtosecond-duration pulsed fields
and, in principle, allow for conducting all-optical setup experiments~\cite{Phuoc}. Note that the radiation produced by either technique is very bright, tunable, 
nearly monoenergetic, and well collimated. These unique features make for a plethora of applications of the generated
MeV radiation, including applications in natural sciences and medicine (for more details see, for instance,~\cite{Umstadter,review1,review2}). 
Here, let us only mention that these radiation sources are the main experimental tool for nuclear physics and astrophysics research 
soon to be performed at ELI~\cite{ELI}.

The key idea for generating the MeV radiation in the aforementioned setups is to allow relativistic electrons move in an intense laser pulse, which forces them to oscillate and radiate; 
the mechanism known as {\it Thomson scattering} in the classical domain, with its quantum generalization known as {\it Compton scattering}. 
Consider a high-energy electron beam colliding with an intense pulse. When moving against the pulse, the electrons experience a nearly flat wave front
of the pulse. Thus, it is physically justified to describe the driving field as a pulsed plane wave~\cite{Neville}. This approach has been
used recently in connection to not only Thomson and Compton scattering~\cite{Boca2009,Boca2011,Boca2012,KKcompton,Mack,Seipt2011,KKpol,KKscale,Macken,Narozhny1,KKscale,Mackenroth}, 
but also to other strong-field processes such as Bethe-Heitler pair creation~\cite{KMK2013}, bremsstrahlung~\cite{Rash1,Rash2}, 
and Mott scattering~\cite{Boca2013,Rash3}. We will use this model in the present paper when investigating the possibility of zepto- or even yoctosecond pulse generation 
from Thomson and Compton processes.

It is known that Thomson and Compton scattering by a finite laser pulse lead to a spectral broadening and a modulation of emitted 
radiation, which is due to spectral interferences from a ramp-on and a ramp-off parts of the driving pulse (cf., Refs.~\cite{Boca2009,Boca2011,KKcompton}). 
Note that, in contrast, a very recent classical calculation by Ghebregziabher {\it et. al.}~\cite{Ghebregziabher}
showed that by chirping a driving pulse, the spectral broadening of high-energy radiation can be reduced. In this paper,
which essentially deals with both Thomson and Compton processes, we observe additional structures of emitted radiation.
We also demonstrate an appearance of a {\it supercontinuum} which arises from spectral broadening.

It is commonly understood that the supercontinuum is a broad bandwidth radiation generated by an interaction 
of a narrow bandwidth laser beam with matter. Moreover, such a spectrum should be spatially and/or temporally coherent. 
In this paper, we investigate coherence properties of radiation generated by nonlinear 
Thomson (Compton) scattering. We also show that the Thomson-originated (Compton-originated) 
supercontinuum allows one to produce {\it zeptosecond} (possibly even {\it yoctosecond}) pulses.

Currently, the shortest optical pulses produced in a laboratory last for 67 attoseconds~\cite{Zhang}. Attosecond pulses 
are routinely produced via high-order harmonic generation (HHG)~\cite{HHG11,HHG22,Farkas} (for recent developments, see also Ref.~\cite{Keitel}). In order to decrease the pulse duration, 
significant progress has to be made to allow for the generation of ultra-high-order harmonics. Specifically, this can be achieved 
using midinfrared driving laser fields (see, e.g., Refs.~\cite{Shan,Jaron}). Alternatively, x-ray radiation from free electron lasers~\cite{FEL} can serve as a novel tool 
for synthesis of ultra-short pulses, which are expected to reach hundreds of zeptoseconds~\cite{Dunning}.
In the present paper, we show a possibility of generating sub-zeptosecond pulses.

The paper is organized as follows. For the convenience of the reader, in Sec.~\ref{theory}, we repeat the theoretical formulation
of Thomson and Compton scattering as introduced in our previous papers~\cite{KKcompton,KKscale}. In Sec.~\ref{amplitude},
we present and compare the frequency spectra produced in both processes. Our further analysis is performed for such parameters
that both spectra coincide. In fact, actual numerical results presented in Secs.~\ref{super} and~\ref{power} relate to Thomson 
classical theory. In Sec.~\ref{super}, we observe the appearance of broad bandwidth radiation, with an energy spread of a few MeV.
The time-dependence of this radiation is investigated in Sec.~\ref{power}. Finally, Sec.~\ref{conclusions} summarizes our results
and draws the conclusions which follow from our study.

\section{Theory}
\label{theory}

The purpose of this section is to introduce notation and to present key formulas for Compton and Thomson scattering spectra. 
For details of the quantum and classical formulation regarding each process we refer the reader to Refs.~\cite{KKcompton,KKpol,KKscale}
and~\cite{KKscale,Jackson1975,LL2,Hart,Hartem,Salamin}, respectively. Since the nonlinear Compton scattering is a direct generalization 
of the nonlinear Thomson scattering into the quantum domain, we shall compile theory for these two processes such that the corresponding 
formulas are as similar as possible.

Throughout the paper, we keep $\hbar=1$. Hence, the fine-structure constant equals $\alpha=e^2/(4\pi\varepsilon_0c)$. 
We use this constant in expressions derived from classical electrodynamics as well, where it is meant to be 
multiplied by $\hbar$ in order to restore the physical units.

\subsection{Basic notation}

Our aim is to define the frequency-angular distribution of the electromagnetic energy that is emitted during either Compton or Thomson scattering 
in the form of outgoing spherical waves. Their polarization is given by a complex unit vector $\bm{\varepsilon}_{\bm{K}\sigma}$, 
where $\sigma=\pm$ labels two polarization degrees of freedom, and where $\bm{K}$ is the wave vector of  radiation emitted in the direction 
$\bm{n}_{\bm{K}}$. Note that ${\bm K}$ determines also the frequency of the emitted radiation since $\omega_{\bm{K}}=c|\bm{K}|$. The wave four-vector, $K$, is therefore 
$K=(\omega_{\bm{K}}/c)(1,\bm{n}_{\bm{K}})$ where $K^2=0$ and $K\cdot\varepsilon_{\bm{K}\sigma}=0$ (we keep $\varepsilon_{\bm{K}\sigma}=(0,\bm{\varepsilon}_{\bm{K}\sigma})$ 
and $\varepsilon_{\bm{K}\sigma}\cdot\varepsilon_{\bm{K}\sigma'}^*=-\delta_{\sigma\sigma'}$). We also assume that three vectors, $(\bm{\varepsilon}_{\bm{K}+},\bm{\varepsilon}_{\bm{K}-},\bm{n}_{\bm{K}})$ 
form the right-handed system of mutually orthogonal unit vectors such that $\bm{\varepsilon}_{\bm{K}+}\times\bm{\varepsilon}_{\bm{K}-}=\bm{n}_{\bm{K}}$.

The laser field which drives both the nonlinear Compton and Thomson scattering is modeled as 
a linearly polarized, pulsed plane wave field, with the following vector potential,
\begin{equation}
\bm{A}(\phi)=A_0 B\bm{\varepsilon}f(\phi).
\label{t1}
\end{equation}
Here, the real vector $\bm{\varepsilon}$ determines the linear polarization of the pulse, 
and the shape function $f(\phi)$ is defined via its derivative
\begin{equation}
f'(\phi)=\begin{cases} 0, & \phi <0, \cr
                 \sin^2\bigl(\frac{\phi}{2}\bigr)\sin(N_{\mathrm{osc}}\phi), & 0\leqslant\phi\leqslant 2\pi,\cr
								0, & \phi > 2\pi,
			  \end{cases}
\label{t2}
\end{equation}
where we assume that $f(0)=0$. Above, $N_{\mathrm{osc}}$ determines the number of cycles in the pulse.

\begin{figure}
\includegraphics[width=7.5cm]{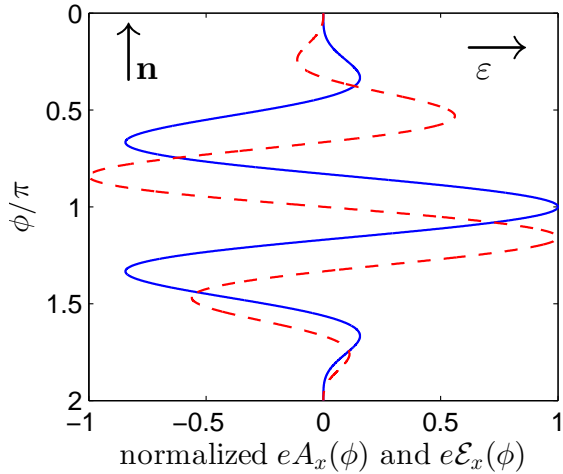}%
\caption{(Color online)  The $\phi$-dependence of the vector potential (solid line), Eq.~\eqref{t1}, and the electric field (dashed line), 
Eq.~\eqref{t6}, multiplied by the electron charge and normalized to their maximum values are plotted for $N_{\mathrm{osc}}=3$. The pulse 
propagates in the $z$-direction, $\bm{n}=\bm{e}_z$, and the linear polarization vector points into the $x$-direction, $\bm{\varepsilon}=\bm{e}_x$. 
The vector potential curve possesses the mirror symmetry with respect to the horizontal line $\phi/\pi=1$, and the electric field line exhibits the axial symmetry with respect to the point $(0,1)$.
\label{polafig20130819}}
\end{figure}

Let us further assume that the duration of the laser pulse is $T_{\mathrm{p}}$. This allows us to introduce 
the fundamental, $\omega=2\pi/T_{\mathrm{p}}$, and the central frequency, $\omega_{\mathrm{L}}=N_{\mathrm{osc}}\omega$,
of the laser field. Moreover, if the laser-pulse propagates in a direction given by the unit vector $\bm{n}$, 
we can define the laser-field four-vector $k=(\omega/c)(1,\bm{n})$ such that $k^2=0$. Hence, the phase $\phi$ in Eq.~\eqref{t1} becomes
\begin{equation}
\phi=k\cdot x=\omega\Bigl(t-\frac{\bm{n}\cdot\bm{r}}{c}\Bigr).
\label{t4}
\end{equation}
For our further purposes we introduce the dimensionless and relativistically invariant parameter
\begin{equation}
\mu=\frac{|e|A_0}{m_{\mathrm{e}}c},
\label{t5}
\end{equation}
where $e=-|e|$ and $m_{\mathrm{e}}$ are the electron charge and the electron rest mass, respectively. With these notations, 
the laser electric and magnetic fields are equal to
\begin{equation}
\bm{\mathcal{E}}(\phi)=\frac{\omega m_{\mathrm{e}}c\mu}{e}B\bm{\varepsilon}f'(\phi),
\label{t6}
\end{equation}
and
\begin{equation}
\bm{\mathcal{B}}(\phi)=\frac{\omega m_{\mathrm{e}}\mu}{e}B(\bm{n}\times\bm{\varepsilon})f'(\phi).
\label{t7}
\end{equation}
The vector potential, Eq.~\eqref{t1}, and the electric component of the laser pulse, Eq.~\eqref{t6}, 
normalized to their maximum values, are presented in Fig.~\ref{polafig20130819} as functions of $\phi$.

In what follows, we put $B=N_{\mathrm{osc}}$, as with this choice and for a given central laser-field frequency $\omega_{\mathrm{L}}$, 
the averaged intensity $I$ of the laser pulse is independent of $N_{\mathrm{osc}}$ \cite{KMK2013}. Specifically,
\begin{equation}
I=A_{\mathrm{I}}\Bigl(\frac{\omega_{\mathrm{L}}}{m_{\mathrm{e}}c^2}\Bigr)^2\mu^2\langle f'^2\rangle ,
\label{t8}
\end{equation}
where
\begin{equation}
\langle f'^2\rangle =\frac{1}{2\pi}\int_0^{2\pi} f'^2(\phi)\mathrm{d}\phi 
\label{t3}
\end{equation}
is equal to $3/16$ for $N_{\mathrm{osc}}>1$, and $5/32$ for $N_{\mathrm{osc}}=1$. If the intensity $I$ is measured in the units of W/cm$^2$, then $A_{\mathrm{I}}=4.6\times 10^{29}$.

\subsection{Energy distributions}

The theory of nonlinear Compton scattering induced by an intense laser pulse was presented in detail 
in Ref. \cite{KKcompton}. It follows from there that the frequency-angular distribution 
of radiated Compton energy (Eqs.~(51) and (52) in Ref. \cite{KKcompton}) can be written down as
\begin{equation}
\frac{\mathrm{d}^3E_{\mathrm{C}}(\bm{K},\sigma;\bm{p}_{\mathrm{i}},\lambda_{\mathrm{i}};\bm{p}_{\mathrm{f}},\lambda_{\mathrm{f}})}{\mathrm{d}\omega_{\bm{K}}\mathrm{d}^2\Omega_{\bm{K}}}
=\alpha |\mathcal{A}_{\mathrm{C},\sigma}(\omega_{\bm{K}},\lambda_{\mathrm{i}},\lambda_{\mathrm{f}})|^2.
\label{t9}
\end{equation}
The above formula relates to the electromagnetic energy emitted (as spherical outgoing waves) 
if the initial electron has momentum $\bm{p}_{\mathrm{i}}$ and spin polarization $\lambda_{\mathrm{i}}$, whereas 
the final electron has the spin polarization $\lambda_{\mathrm{f}}$ and its momentum is determined by the momentum 
conservation equations (cf., Eqs. (47) in Ref. \cite{KKcompton}). We shall call the complex function 
$\mathcal{A}_{\mathrm{C},\sigma}(\omega_{\bm{K}},\lambda_{\mathrm{i}},\lambda_{\mathrm{f}})$ the Compton amplitude; 
it is worth noting that it also depends on the remaining parameters of the Compton scattering, but we display 
explicitly only those which are essential for our further discussion. If one is not interested in the dependence 
of emitted radiation on the electron spin degrees of freedom, than the distribution above has to be summed over the final 
and averaged over the initial spin polarizations. This leads to 
\begin{equation}
\frac{\mathrm{d}^3E_{\mathrm{C}}(\bm{K},\sigma)}{\mathrm{d}\omega_{\bm{K}}\mathrm{d}^2\Omega_{\bm{K}}}
=\frac{\alpha}{2} \sum_{\lambda_{\mathrm{i}},\lambda_{\mathrm{f}}=\pm}|\mathcal{A}_{\mathrm{C},\sigma}(\omega_{\bm{K}},\lambda_{\mathrm{i}},\lambda_{\mathrm{f}})|^2 ,
\label{t10}
\end{equation}
where all non-relevant electron degrees of freedom are hidden.

The complete theory of nonlinear Thomson scattering is presented in Jackson's textbook~\cite{Jackson1975} (see, also Ref. \cite{KKscale}). 
The relevant frequency-angular distribution of energy emitted during this process can be expressed as
\begin{equation}
\frac{\mathrm{d}^3E_{\mathrm{Th}}(\bm{K},\sigma)}{\mathrm{d}\omega_{\bm{K}}\mathrm{d}^2\Omega_{\bm{K}}}=\alpha|\mathcal{A}_{\mathrm{Th},\sigma}(\omega_{\bm{K}})|^2,
\label{t11}
\end{equation}
and it should be compared with Eqs.~\eqref{t9} or \eqref{t10} for Compton scattering. In analogy with the Compton theory, the complex function 
$\mathcal{A}_{\mathrm{Th},\sigma}(\omega_{\bm{K}})$ will be called here the Thomson amplitude. As shown in Ref. \cite{KKscale}, its explicit form can be represented as an integral 
\begin{equation}
\mathcal{A}_{\mathrm{Th},\sigma}(\omega_{\bm{K}})=\frac{1}{2\pi}\int_0^{2\pi}\mathrm{d}\phi \Upsilon_{\sigma}(\phi)\exp(\mathrm{i}\omega_{\bm{K}}\ell(\phi)/c),
\label{t12}
\end{equation}
where
\begin{eqnarray}
\ell(\phi)&=&c\frac{\phi}{\omega}+(\bm{n}-\bm{n}_{\bm{K}})\cdot \bm{r}(\phi),\label{t13}\\
\Upsilon_{\sigma}(\phi)&=&\bm{\varepsilon}_{\bm{K}\sigma}^*\cdot \frac{\bm{n}_{\bm{K}}\times [(\bm{n}_{\bm{K}}-\bm{\beta}(\phi))\times \bm{\beta}'(\phi)]}{\bigl(1-\bm{n}_{\bm{K}}\cdot \bm{\beta}(\phi)\bigr)^2},
\label{t14}
\end{eqnarray}
and where the electron position $\bm{r}(\phi)$ and reduced velocity $\bm{\beta}(\phi)$ fulfill the system of ordinary differential equations
\begin{align}
\frac{\mathrm{d}\bm{r}(\phi)}{\mathrm{d}\phi}=&\frac{c}{\omega}\frac{\bm{\beta}(\phi)}{1-\bm{n}\cdot\bm{\beta}(\phi)}, \label{thom9ex} \\
\frac{\mathrm{d}\bm{\beta}(\phi)}{\mathrm{d}\phi}=&\mu\frac{\sqrt{1-\bm{\beta}^2(\phi)}}{1-\bm{n}\cdot\bm{\beta}(\phi)} \nonumber \\
\times & \Bigl[ 
\bigl(\bm{\varepsilon}-\bm{\beta}(\phi)(\bm{\beta}(\phi)\cdot\bm{\varepsilon})+\bm{\beta}(\phi)\times(\bm{n}\times\bm{\varepsilon})\bigr)f^{\prime}(\phi) \Bigr] .\nonumber
\end{align}
Let us recall that these equations can be derived from the Newton-Lorentz relativistic equations, with time $t$ which relates to the phase $\phi$ by Eq.~\eqref{t4}.

\begin{figure}
\includegraphics[width=7.5cm]{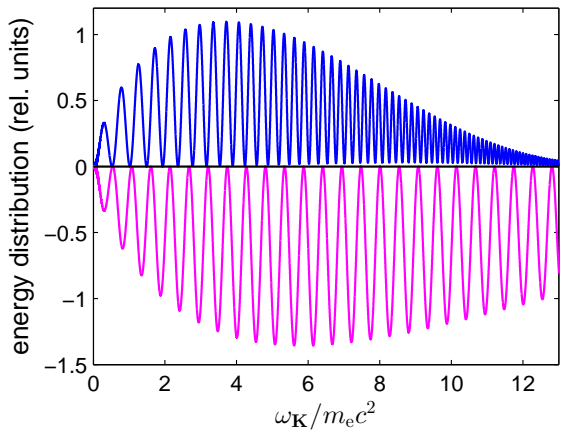}%
\caption{(Color online) Energy spectra for Compton scattering (upper panel), Eq.~\eqref{t10}, and for Thomson scattering 
(lower panel, reflected with respect to the horizontal black line), Eq.~\eqref{t11}. The laser field parameters are such that 
$\mu=10$, $N_{\mathrm{osc}}=3$, $\omega_{\mathrm{L}}=0.03 m_{\mathrm{e}}c^2$, and the scattered radiation is linearly polarized 
in the scattering plane, i.e., in the $(xz)$-plane. The direction of scattered radiation is given by the polar and azimuthal angles, 
$\theta_{\bm{K}}=0.1\pi$ and $\varphi_{\bm{K}}=\pi$, respectively. These parameters are specified in the rest frame of incident electrons.
\label{sc3b20130820}}
\end{figure}

\section{Scattering amplitudes}
\label{amplitude}

\begin{figure}
\includegraphics[width=7.5cm]{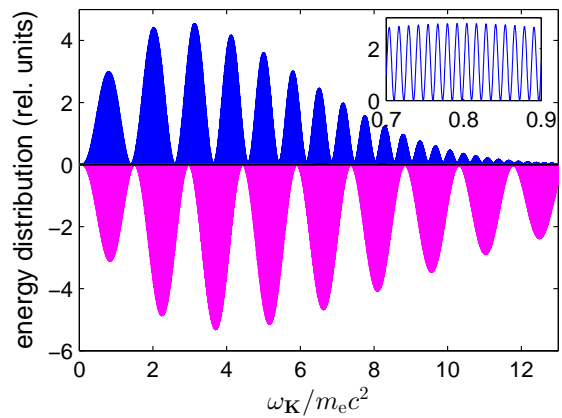}%
\caption{(Color online) The same as in Fig.~\ref{sc3b20130820} but for the polar and azimuthal angles, $\theta_{\bm{K}}=0.1\pi$ and $\varphi_{\bm{K}}=0$,
respectively. In the inset, we present the enlarged portion of the Compton distribution with regular oscillations of the energy spectrum.
\label{sc3ainsmod20130820}}
\end{figure}

Let us start by comparing the quantum Compton process with its classical approximation which is Thomson process, both driven by a three-cycle laser pulse ($N_{\rm osc}=3$)
with the electric and magnetic fields defined by the shape function \eqref{t2} (see, also Fig.~\ref{polafig20130819}). 
We choose the reference frame such that the pulse propagates in the $z$-direction, $\bm{n}=\bm{e}_z$, its linear polarization 
vector is $\bm{\varepsilon}=\bm{e}_x$, whereas the initial electron velocity vanishes, i.e., initially electrons are at rest. 
In Figs.~\ref{sc3b20130820} and~\ref{sc3ainsmod20130820} we compare the quantum and classical theories for the same pulse configuration 
but for two different directions of observation of scattered radiation. For both directions of emission, we see that the Compton spectrum 
is compressed in comparison with the Thomson spectrum. As discussed in Ref. \cite{KKscale}, such a nonlinear compression of the Compton distribution 
(note that the larger the frequency the more compressed the spectrum is) indicates the quantum nature of Compton scattering.
If for the Thomson scattering we denote the frequency as $\omega^{\mathrm{Th}}_{\bm{K}}$ and scale it to $\omega_{\bm{K}}$ according to the rules
\begin{equation}
\omega_{\mathrm{cut}}=c\frac{n\cdot p_{\mathrm{i}}}{n\cdot n_{\bm{K}}}=\frac{m_{\mathrm{e}}c^2}{1-\cos\theta_{\bm{K}}}\approx 20.4m_{\mathrm{e}}c^2 ,
\label{fre4}
\end{equation}
and
\begin{equation}
\omega_{\bm{K}}^{\mathrm{Th}}-\omega_{\bm{K}}=\frac{\omega_{\bm{K}}^{\mathrm{Th}}\omega_{\bm{K}}}{\omega_{\mathrm{cut}}} ,
\label{fre6}
\end{equation}
then both spectra become similar to each other. Namely, they have maxima and minima for the same frequencies, but their absolute values 
can differ. In Eq.~\eqref{fre4}, we have introduced the cut-off frequency, $\omega_{\mathrm{cut}}$, which defines the maximum frequency 
for the Compton scattering. It appears that both quantum and classical approaches give the same results for the aforementioned absolute values,
provided that $\omega_{\bm{K}}\ll \omega_{\mathrm{cut}}$. Moreover, as it follows from the discussion presented in Ref. \cite{KKscale}, 
even for frequencies close to the cut-off both distributions remain similar when helicities of the initial and final electron
states are the same.

However, when comparing Figs. \ref{sc3b20130820} and \ref{sc3ainsmod20130820}, we observe that presented distributions 
look qualitatively different. While for the first case we observe slowly pulsating spectra, for the second case 
the spectra are modulated and exhibit very rapid oscillations. It is well-known that the reason for the oscillatory 
behavior of energy distributions is the interference of scattered radiation, but this does not explain the qualitative 
difference between these two cases. It could be attributed, for instance, to the right-left asymmetry of the laser field potential 
(Fig. \ref{polafig20130819}), as it has been discussed for the case of electron-positron pair creation~\cite{KKasymm}. 
In order to analyze this problem in detail we found it very difficult (if not impossible) to dwell on the Compton theory, 
due to its complicated character. However, such a discussion can be based on the Thomson theory, which we present below. 
We stress that electron-laser-field scattering with emission of extra photons is actually described 
by the QED Compton scattering, and that Thomson scattering can be {\it only} treated as its approximation. Therefore, the classical theory 
can be used to interpret the results whenever (within the range of its applicability) it is difficult to provide a reasonable physical interpretation
based on the more complete quantum theory.

\begin{figure}
\includegraphics[width=7.5cm]{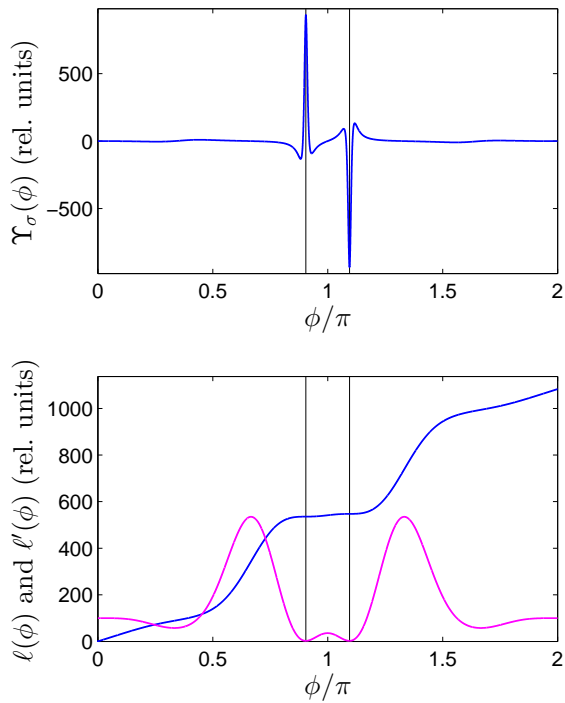}%
\caption{(Color online) The functions $\Upsilon_{\sigma}(\phi)$, Eq.~\eqref{t14}, (upper frame), and $\ell(\phi)$ and its derivative 
$\ell'(\phi)$, Eq.~\eqref{t13}, (lower frame) for the scattering parameters of Fig. \ref{sc3b20130820}. Two thin vertical lines mark 
the positions of the dominant extrema of $\Upsilon_{\sigma}(\phi)$, for $\phi_1\approx 0.9\pi$ and $\phi_2=2\pi-\phi_1$. In the lower frame,
the blue (dark and monotonously increasing) line represent the $\ell(\phi)$ function whereas the magenta (gray and exhibiting extrema) line the $\ell'(\phi)$ function.
\label{classic0d20130819}}
\end{figure}

Let us consider the first case, Fig. \ref{sc3b20130820}, and draw the functions $\Upsilon_{\sigma}(\phi)$, $\ell(\phi)$, 
and its derivative $\ell'(\phi)$, as presented in Fig. \ref{classic0d20130819}. We observe here two main extrema of $\Upsilon_{\sigma}(\phi)$ 
for $\phi_1\approx 0.9\pi$ and $\phi_2=2\pi-\phi_1$, the position of which coincide with two global minima of $\ell'(\phi)$. Since
\begin{equation}
\ell'(\phi)=\frac{c}{\omega}\frac{1-\bm{n}_{\bm{K}}\cdot\bm{\beta}(\phi)}{1-\bm{n}\cdot\bm{\beta}(\phi)} > 0,
\label{t15}
\end{equation}
we attribute these extrema to the minimum value of the denominator in Eq.~\eqref{t14}. Moreover, the function $\Upsilon_{\sigma}(\phi)$ 
exhibits a very sharp maximum and minimum, therefore, the Thomson amplitude, Eq.~\eqref{t12}, can be approximated by two terms
\begin{equation}
\mathcal{A}_{\mathrm{Th},\sigma}(\omega_{\bm{K}})\approx A(\mathrm{e}^{\mathrm{i}\omega_{\bm{K}}\ell(\phi_1)/c}-\mathrm{e}^{\mathrm{i}\omega_{\bm{K}}\ell(\phi_2)/c}).
\label{t16}
\end{equation}
Here, $A$ is roughly equal to the area under the peak. More precisely, $A$ is a slowly varying function of $\omega_{\bm K}$ which 
can be calculated by applying a better approximation, for instance, the saddle point method for sufficiently large $\omega_{\bm K}$. 
However, for the purpose of our further discussion it is sufficient to consider $A$ as a constant. 

It follows from the numerical data that
\begin{equation}
\ell(\phi_1)=\tau_0-\tau_{\mathrm{small}}\quad\mathrm{and}\quad \ell(\phi_2)=\tau_0+\tau_{\mathrm{small}},
\label{t17}
\end{equation}
with $\tau_0\approx 541.5/m_{\mathrm{e}}c$ and $\tau_{\mathrm{small}}\approx 5.9/m_{\mathrm{e}}c$. The modulus squared of the pulsating part of Thomson amplitude is 
therefore proportional to
\begin{equation}
|\mathcal{A}_{\mathrm{Th},\sigma}(\omega_{\bm{K}})|^2\propto \sin^2\Bigl(5.9\frac{\omega_{\bm{K}}}{m_{\mathrm{e}}c^2}\Bigr).
\label{t18}
\end{equation}
This means that two consecutive minima in the Thomson distribution are separated by $\Delta\omega_{\bm{K}}/ m_{\mathrm{e}}c^2\approx \pi/5.9\approx 0.53$, 
which agrees very well with the data presented in Fig. \ref{sc3b20130820}.

\begin{figure}
\includegraphics[width=7.5cm]{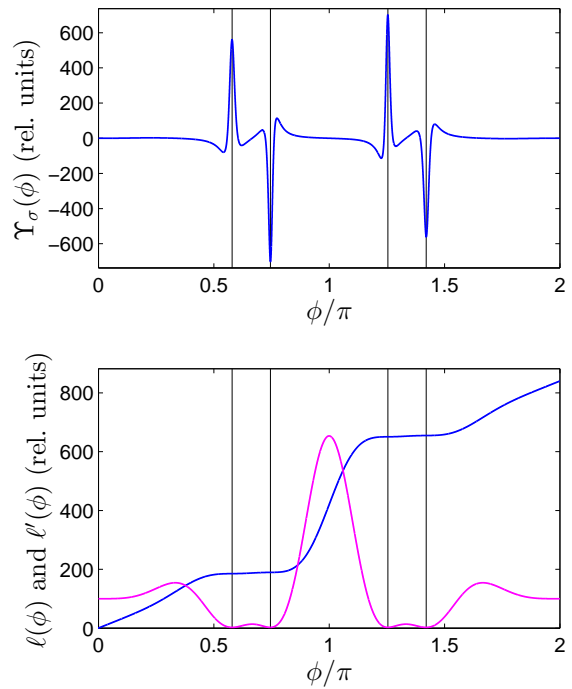}%
\caption{(Color online) The same as in Fig. \ref{classic0d20130819} but for the scattering parameters of Fig. \ref{sc3ainsmod20130820}. 
Four thin vertical lines indicate positions of the dominant extrema of $\Upsilon_{\sigma}(\phi)$, for $\phi_1\approx 0.58\pi$, $\phi_2=0.75\pi$, 
$\phi_3=2\pi-\phi_1$, and $\phi_4=2\pi-\phi_2$. The analysis of the numerical data shows that $\ell(\phi_1)=\tau_0-\tau_{\mathrm{big}}-\tau_{\mathrm{small}}$, 
$\ell(\phi_2)=\tau_0-\tau_{\mathrm{big}}+\tau_{\mathrm{small}}$, $\ell(\phi_3)=\tau_0+\tau_{\mathrm{big}}-\tau_{\mathrm{small}}$, and 
$\ell(\phi_4)=\tau_0+\tau_{\mathrm{big}}+\tau_{\mathrm{small}}$, with $\tau_0 \approx 420/m_{\mathrm{e}}c$, $\tau_{\mathrm{big}} \approx 233/m_{\mathrm{e}}c$, 
and $\tau_{\mathrm{small}} \approx 2.14/m_{\mathrm{e}}c$. 
\label{classic1d20130819}}
\end{figure}

A similar interpretation can be attributed to the modulated and rapidly oscillating energy distributions presented in Fig. \ref{sc3ainsmod20130820}. 
In this case we have four main extrema (cf., Fig. \ref{classic1d20130819}). Although the outer extrema are smaller in magnitude than the inner ones, 
they are also more spread out. Therefore, in the first approximation, we can assume that the related areas (under the maxima and above the minima) are equal to each other.
This leads to the approximate form for the Thomson amplitude,
\begin{equation}
\mathcal{A}_{\mathrm{Th},\sigma}(\omega_{\bm{K}})\approx A\sum_{s=1}^4 (-1)^{s+1}\mathrm{e}^{\mathrm{i}\omega_{\bm{K}}\ell(\phi_s)/c}.
\label{t19}
\end{equation}
Hence, 
\begin{equation}
|\mathcal{A}_{\mathrm{Th},\sigma}(\omega_{\bm{K}})|^2\propto \sin^2\Bigl(2.14\frac{\omega_{\bm{K}}}{m_{\mathrm{e}}c^2}\Bigr)\sin^2\Bigl(233\frac{\omega_{\bm{K}}}{m_{\mathrm{e}}c^2}\Bigr),
\label{t20}
\end{equation}
which again agrees very well with the data presented in Fig. \ref{sc3ainsmod20130820}. In this case, the consecutive minima of fast oscillations 
are separated by $\Delta\omega_{\bm{K}}/ m_{\mathrm{e}}c^2\approx \pi/233\approx 0.013$, whereas for the modulation we obtain $\delta\omega_{\bm{K}}/ m_{\mathrm{e}}c^2\approx \pi/2.14\approx 1.46$.
In other words, there are more than 100 oscillations within a single modulation.

\section{Generation of supercontinuum}
\label{super}

Since its demonstration in the early 1970s \cite{super1}, the supercontinuum generation has been the focus of significant research activities. 
It has attracted much attention owing to its enormous spectral broadening (for instance, it is possible to obtain a white light spectrum 
covering the entire visible range from 400 to 700 nm), which have many useful applications in telecommunication \cite{super2}, frequency 
metrology \cite{super3}, optical coherence tomography \cite{super4}, and device characterization \cite{super5}. It has to be mentioned, 
however, that the supercontinuum generation in photonics is, in general, a complex physical phenomenon involving many nonlinear optical 
effects such as self-phase modulation, cross phase modulation, four wave mixing, and stimulated Raman scattering \cite{super6}. 
It seems to be simpler to analyze the generation of a broadband spectrum of radiation by the Thomson and Compton scattering.

A closer look at Fig.~\ref{sc3b20130820} 
suggests that the Thomson (or Compton) process could be used for the generation of a supercontinuum with the bandwidth of few keV and with a small change 
of its intensity. It is shown, for instance, by the first pulsation in the energy distribution, the width of which is around 200keV in 
the reference frame of electrons. In order to further investigate such a possibility we study below nonlinear Thomson scattering in the laboratory frame. 
Since we are going to consider the spectrum of frequencies much smaller than the cut-off frequency \eqref{fre4}, 
the Thomson and Compton theories give practically the same results. The former, however, can be treated numerically much faster.

In the analysis of Thomson and Compton processes in the laboratory frame we have to account for the fact that the initial energy of electrons 
has to be large, as compared to $m_{\mathrm{e}}c^2$, in order to generate sufficiently intense pulses of scattered radiation. Moreover, the laser 
pulse central frequency $\omega_{\mathrm{L}}$ is much smaller then $m_{\mathrm{e}}c^2$. This means that majority of the generated radiation is 
emitted in a very sharp cone. In particular for a head-on collision of the laser and electron beams, when electrons are moving in the direction 
opposite to the $z$ axis, the radiation is scattered for $\theta_{\bm{K}}$ very close to $\pi$. In this case the parametrization of all possible 
directions of emission by two spherical angles $\theta_{\bm{K}}$ and $\varphi_{\bm{K}}$ is not convenient, as it will follow shortly. 
It is better to consider another system of Cartesian coordinates $(x',y',z')$ such that (cf. Refs.~\cite{KKcompton,KKbreit})
\begin{equation}
(x',y',z')=(z,x,y).
\label{t23}
\end{equation}
If, in the new system of coordinates, we denote the polar angle by $\Phi_{\bm{K}}$ ($0\leqslant \Phi_{\bm{K}}\leqslant \pi$) and 
the azimuthal angle by $\Theta_{\bm{K}}$ ($0\leqslant\Theta_{\bm{K}}<2\pi$), then they can be related to the original polar and 
azimuthal angles $\theta_{\bm{K}}$ and $\varphi_{\bm{K}}$ by the equations:
\begin{equation}
\Phi_{\bm{K}}=\arccos(\sin\theta_{\bm{K}}\sin\varphi_{\bm{K}}) 
\label{t24}
\end{equation}
and
\begin{equation}
\tan\Theta_{\bm{K}}=\tan\theta_{\bm{K}}\cos\varphi_{\bm{K}}.
\label{t25}
\end{equation}
In addition, the scattering plane $(xz)$ which before was defined by two conditions, $\varphi_{\bm{K}}=0$ and $\varphi_{\bm{K}}=\pi$, 
now is defined by the single condition, $\Phi_{\bm{K}}=\pi/2$. The discontinuous change of angles 
$(\theta_{\bm{K}},\varphi_{\bm{K}})=(\pi-\varepsilon,0)\rightarrow (\pi-\varepsilon,\pi)$, $\varepsilon \ll 1$, 
which corresponds to the continuous change of directions near the south pole of the unit sphere, now is described by 
the continuous change of angles $(\Phi_{\bm{K}},\Theta_{\bm{K}})=(\pi/2,\pi-\varepsilon)\rightarrow (\pi/2,\pi+\varepsilon)$. 
It is, therefore, more convenient to use the angles $(\Phi_{\bm{K}},\Theta_{\bm{K}})$ in our analysis. Since the infinitesimal solid angle becomes
\begin{equation}
\mathrm{d}^2\Omega_{\bm{K}}=\sin\Phi_{\bm{K}}\mathrm{d}\Phi_{\bm{K}}\mathrm{d}\Theta_{\bm{K}},
\label{t26}
\end{equation}
we can define the partially integrated energy spectrum of emitted radiation for the Thomson process
\begin{equation}
\frac{\mathrm{d}^2E_{\mathrm{Th}}(\bm{K},\sigma)}{\sin\Phi_{\bm{K}}\mathrm{d}\omega_{\bm{K}}\mathrm{d}\Phi_{\bm{K}}}=\int_0^{2\pi}\mathrm{d}\Theta_{\bm{K}}\ \frac{\mathrm{d}^3E_{\mathrm{Th}}(\bm{K},\sigma)}{\mathrm{d}\omega_{\bm{K}}\mathrm{d}^2\Omega_{\bm{K}}},
\label{t27}
\end{equation}
and similarly for the Compton process. Next, we can define the angular distribution,
\begin{equation}
\frac{\mathrm{d}^2E_{\mathrm{Th}}(\bm{n}_{\bm{K}},\sigma)}{\mathrm{d}^2\Omega_{\bm{K}}}=\int_0^{\omega_{\mathrm{max}}}\mathrm{d}\omega_{\bm{K}}\ \frac{\mathrm{d}^3E_{\mathrm{Th}}(\bm{K},\sigma)}{\mathrm{d}\omega_{\bm{K}}\mathrm{d}^2\Omega_{\bm{K}}},
\label{t28}
\end{equation}
where we have introduced the maximum frequency, $\omega_{\rm max}$. This
frequency is infinite for Thomson scattering, but it is equal to the cut-off frequency \eqref{fre4} for Compton scattering \cite{KKscale}, 
independently of the incident pulse duration.

\begin{figure}
\includegraphics[width=8.0cm]{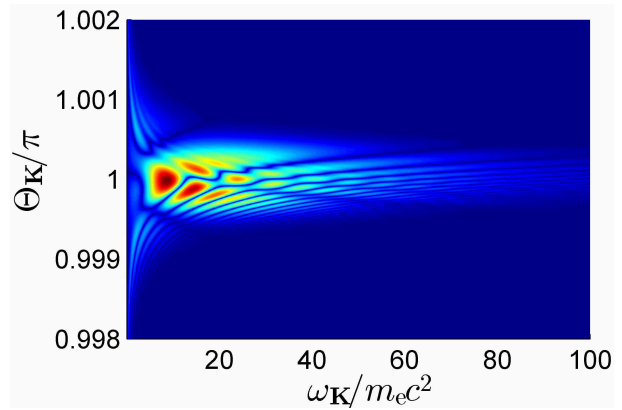}%
\caption{(Color online) Color map of the energy distribution for Thomson process, Eq.~\eqref{t11}. We assume a three-cycle driving pulse, 
linearly polarized in the $x$-direction [for the shape function of the electric field component, see Eq.~\eqref{t2}], counter-propagating
an electron beam. The pulse central frequency in the laboratory frame equals $\omega_{\mathrm{L}}=1.548\mathrm{eV}\approx 3\times 10^{-6}m_{\mathrm{e}}c^2$ 
and its averaged intensity is determined by $\mu=1$ [Eq.~\eqref{t8}]. Electrons move in the opposite $z$-direction, with momentum 
$|\bm{p}_{\mathrm{i}}|=1000m_{\mathrm{e}}c$ and the scattering process occurs in the plane $\Phi_{\bm{K}}=\pi/2$ [see, Eqs.~\eqref{t24} 
and~\eqref{t25} for the definitions of angles $\Phi_{\bm{K}}$ and $\Theta_{\bm{K}}$]. The emitted radiation is linearly polarized in 
the $(xz)$-plane [or, equivalently, in the $(x'y')$-plane].
\label{surf3axrep1d20130904r600}}
\end{figure}

\begin{figure}
\includegraphics[width=8.0cm]{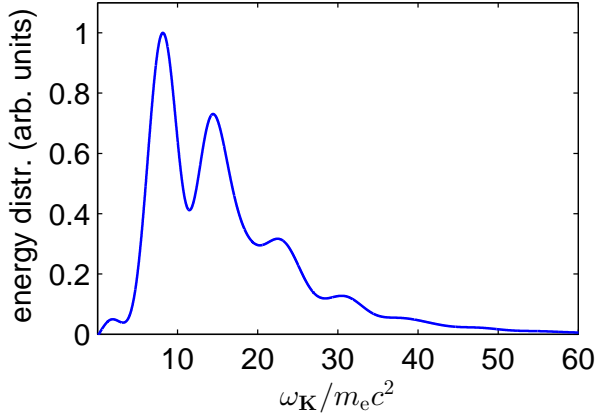}%
\caption{(Color online) The energy differential distribution of emitted radiation integrated over $\Theta_{\bm{K}}$, Eq.~\eqref{t27}, 
and normalized to its maximum value, for the same geometry and parameters as in Fig.~\ref{surf3axrep1d20130904r600}. 
\label{super3axrep1d20130905}}
\end{figure}

In Fig. \ref{surf3axrep1d20130904r600}, we present the color map of energy distribution for the Thomson process as a function 
of frequency and emission angle in the scattering plane, i.e., for $\Phi_{\bm{K}}=\pi/2$. It clearly demonstrates that, for frequencies around 
8$m_{\mathrm{e}}c^2$ (4MeV) and 16$m_{\mathrm{e}}c^2$ (8MeV), intense and very broad (of the order of 2MeV) candidates for the supercontinuum 
are created. Note also that for the considered electron and laser beams parameters, the radiation is scattered within a very narrow cone. Although we have 
presented results for a particular $\Phi_{\bm{K}}$, the pattern preserves its structure also for the polar angle $\Phi_{\bm{K}}$ 
close to $\pi/2$. Fig. \ref{super3axrep1d20130905} shows the partially integrated energy distribution, Eq.~\eqref{t27}, for $\Phi_{\bm{K}}=\pi/2$ 
with two broad peaks. In order to call these structures the supercontinuum we have to investigate their coherent properties. For Compton processes,
this problem has been partially discussed in~\cite{KKcomb}. We have shown there that the phase of the Compton probability amplitude is not random 
and it linearly increases with $\omega_{\bm{K}}$ in the frequency intervals sufficiently wide in order to cover at least a few interference peaks 
in the energy distribution. We have checked that the same occurs for Thomson scattering. This suggests that a synthesis of frequencies from
the supercontinuum indeed can lead to the generation of very short pulses of radiation.

\section{Temporal power distributions}
\label{power}

In the previous section we have demonstrated the possibility of the generation of a very broad spectrum of radiation, with the bandwidth of few MeV. 
In order to show the coherent properties of this spectrum one has to investigate the time-dependence of scattered radiation and to check if it 
can be used for the synthesis of very short pulses. This is the case, for instance, of HHG in gases \cite{HHG}, plasmas \cite{HHG1}, 
and crystals \cite{HHG2}, which are nowadays routinely used for the synthesis of attosecond pulses (for a review concerning HHG, see, also Ref.~\cite{Keitel}). 
Thus, we need to relate the frequency-angular distributions of energy of generated radiation to the temporal power distribution of the emitted radiation. 
For simplicity, we shall do it for the classical theory.

By analyzing the Li\'enard-Wiechert potentials \cite{Jackson1975,LL2} it is straightforward to relate the Thomson amplitude 
$\mathcal{A}_{\mathrm{Th},\sigma}(\omega_{\bm{K}})$ to the electric field of the scattered radiation. Indeed, the Fourier transform 
of the electric field of polarization $\bm{\varepsilon}_{\bm{K}\sigma}$, in the far radiation zone, has the form of the outgoing spherical wave
\begin{equation}
\tilde{\mathcal{E}}_{\sigma}(\omega_{\bm{K}})=\frac{\mathrm{e}^{\mathrm{i}|\bm{K}|R}}{R}\frac{e}{4\pi\varepsilon_0c}2\pi \mathcal{A}_{\mathrm{Th},\sigma}(\omega_{\bm{K}}),
\label{tt1}
\end{equation}
with its space-time form:
\begin{equation}
\mathcal{E}_{\sigma}\Bigl(t-\frac{R}{c}\Bigr)= \int_{-\infty}^{\infty}\frac{\mathrm{d}\omega}{2\pi}\mathrm{e}^{-\mathrm{i}\omega t}\tilde{\mathcal{E}}_{\sigma}(\omega).
\label{tt2}
\end{equation}
In this notation, we show explicitly the time-dependence of the electric field, whereas the decay $1/R$, being less important for our further discussion, 
is hidden. To make the notation shorter, we introduce the retarded phase
\begin{equation}
\phi_{\mathrm{r}}=\omega_0\Bigl(t-\frac{R}{c}\Bigr),
\label{tt3}
\end{equation}
and rewrite Eq.~\eqref{tt2} as
\begin{equation}
\mathcal{E}_{\sigma}(\phi_{\mathrm{r}})=\frac{e}{4\pi\varepsilon_0cR}\tilde{\mathcal{A}}_{\mathrm{Th},\sigma}(\phi_{\mathrm{r}}),
\label{tt4}
\end{equation}
where
\begin{equation}
\tilde{\mathcal{A}}_{\mathrm{Th},\sigma}(\phi_{\mathrm{r}})=\int_{-\infty}^{\infty}\mathrm{d}\omega \mathcal{A}_{\mathrm{Th},\sigma}(\omega)\mathrm{e}^{-\mathrm{i}\omega\phi_{\mathrm{r}}/\omega_0 } .
\label{tt5}
\end{equation}
In the definition of $\phi_{\mathrm{r}}$, we have introduced an arbitrary frequency $\omega_0$. We shall discuss below which value for this parameter should be chosen.

Since the electric field is real, we have $\mathcal{A}_{\mathrm{Th},\sigma}(-\omega)=\mathcal{A}^*_{\mathrm{Th},\sigma}(\omega)$. Defining
\begin{equation}
\tilde{\mathcal{A}}^{(+)}_{\mathrm{Th},\sigma}(\phi_{\mathrm{r}})=\int_0^{\infty}\mathrm{d}\omega \mathcal{A}_{\mathrm{Th},\sigma}(\omega)\mathrm{e}^{-\mathrm{i}\omega\phi_{\mathrm{r}}/\omega_0 },
\label{tt6}
\end{equation}
we find that
\begin{equation}
\mathcal{E}_{\sigma}(\phi_{\mathrm{r}})=\frac{e}{4\pi\varepsilon_0cR}2\Re \tilde{\mathcal{A}}^{(+)}_{\mathrm{Th},\sigma}(\phi_{\mathrm{r}}),
\label{tt7}
\end{equation}
where the symbol $\Re$ means the real value. Applying the Poynting theorem of classical electrodynamics \cite{Jackson1975,LL2} 
we find that the total energy of scattered radiation transmitted through an infinitesimal surface $R^2\mathrm{d}\Omega_{\bm{K}}$ equals
\begin{equation}
\mathrm{d}^2E_{\mathrm{Th},\sigma}=\mathrm{d}^2\Omega_{\bm{K}}\frac{\alpha}{\pi}\int_{-\infty}^{\infty}\mathrm{d}t \bigl(\Re \tilde{\mathcal{A}}^{(+)}_{\mathrm{Th},\sigma}(\phi_{\mathrm{r}})\bigr)^2.
\label{tt8}
\end{equation}
This allows us to define the angular distribution of temporal power of emitted radiation,
\begin{equation}
\frac{\mathrm{d}^2P_{\mathrm{Th},\sigma}(\phi_{\mathrm{r}})}{\mathrm{d}^2\Omega_{\bm{K}}}=\frac{\alpha}{\pi}\bigl(\Re \tilde{\mathcal{A}}^{(+)}_{\mathrm{Th},\sigma}(\phi_{\mathrm{r}})\bigr)^2.
\label{tt9}
\end{equation}
For the Compton scattering, the corresponding temporal power distribution looks similar, except that the Compton amplitude, 
$\mathcal{A}_{\mathrm{C},\sigma}(\omega_{\bm{K}};\lambda_{\mathrm{i}},\lambda_{\mathrm{f}})$, depends also on the electron spin degrees of freedom.

\begin{figure}
\includegraphics[width=7.5cm]{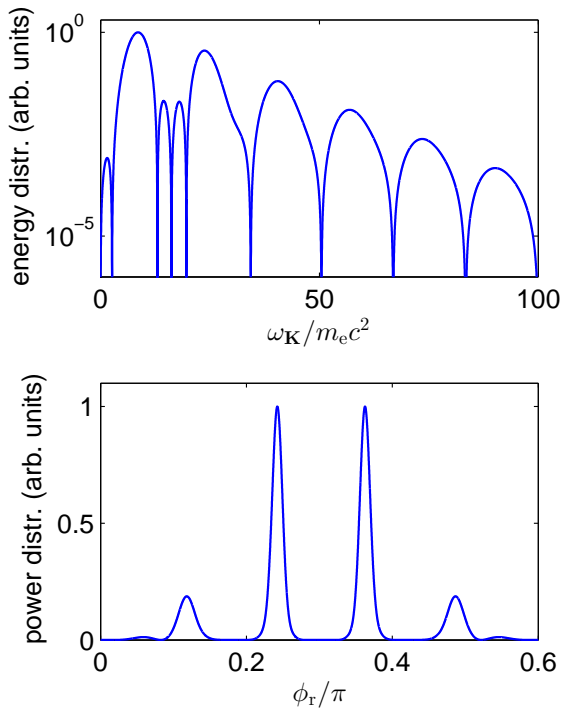}%
\caption{(Color online) Shows the energy distribution (upper panel), Eq.~\eqref{t11}, and the power distribution (lower panel), Eq.~\eqref{tt9}, 
for the Thomson scattering. The geometry and the laser and electron beams parameters are the same as in Fig.~\ref{surf3axrep1d20130904r600}, 
except that the results are presented for the particular $\Theta_{\bm{K}}=1.00001\pi$ [or, equivalently, $(\theta_{\bm{K}},\varphi_{\bm{K}})=(0.99999\pi,\pi)$]. 
The emitted radiation is polarized in the $(xz)$-plane and the distributions are normalized to their maximum values. For the power distribution, 
we put $\omega_0=m_{\mathrm{e}}c^2$ while the upper limit of the frequency integration in~\eqref{tt6} we set to $100m_{\mathrm{e}}c^2$.
\label{puls3axcARd20130907}}
\end{figure}
\begin{figure}
\includegraphics[width=7.5cm]{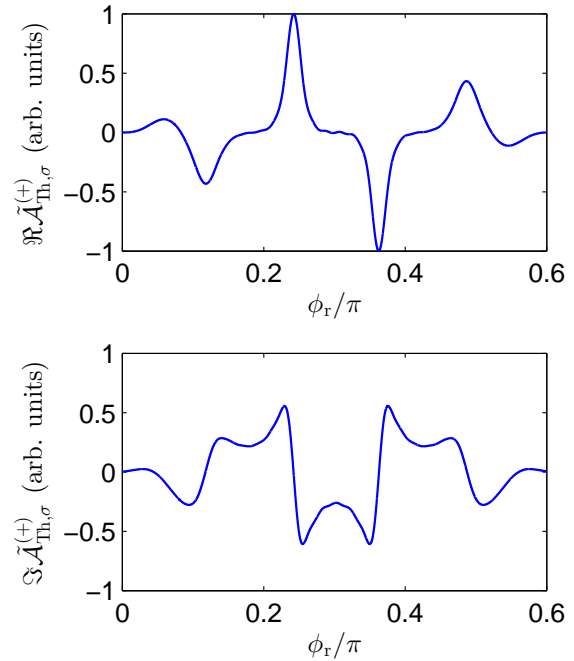}%
\caption{(Color online) The real (upper panel) and imaginary (lower panel) parts of the Fourier transform of the Thomson amplitude,
 Eq.~\eqref{tt6}, normalized to the maximum value of $\Re\tilde{\mathcal{A}}_{\mathrm{Th},\sigma}^{(+)}$. The plots are for the parameters 
from Fig.~\ref{puls3axcARd20130907}. The real part of the Fourier transform is proportional to the electric field of the emitted radiation.
\label{puls3axcBd20130907}}
\end{figure}

Similarly to the integrated energy distributions, Eqs.~\eqref{t27} and \eqref{t28}, we can define the integrated power distribution of Thomson radiation
\begin{equation}
\frac{\mathrm{d}P_{\mathrm{Th},\sigma}(\phi_{\mathrm{r}})}{\sin\Phi_{\bm{K}}\mathrm{d}\Phi_{\bm{K}}}=\int_0^{2\pi} \mathrm{d}\Theta_{\bm{K}}\frac{\mathrm{d}^2P_{\mathrm{Th},\sigma}(\phi_{\mathrm{r}})}{\mathrm{d}^2\Omega_{\bm{K}}},
\label{tt10}
\end{equation}
and its angular distribution
\begin{equation}
\frac{\mathrm{d}^2E_{\mathrm{Th}}(\bm{n}_{\bm{K}},\sigma)}{\mathrm{d}^2\Omega_{\bm{K}}}=\int_0^{\phi_{\mathrm{max}}}\mathrm{d}\phi_{\mathrm{r}}\ \frac{\mathrm{d}^2P_{\mathrm{Th},\sigma}(\phi_{\mathrm{r}})}{\mathrm{d}^2\Omega_{\bm{K}}}.
\label{tt11}
\end{equation}
Here, we have introduced the maximum value $\phi_{\mathrm{max}}$ for the retarded phase $\phi_{\mathrm{r}}$, which has the purely numerical origin and it is closely 
related to the parameter $\omega_0$ used in Eq.~\eqref{tt3}. Indeed, numerically, the Thomson and Compton amplitudes are calculated for some discrete values 
$\omega_{\bm{K}}$; in our case, we choose the equally spaced values with a step $\Delta\omega_{\bm{K}}$. For instance, in our calculations presented in Fig.~\ref{surf3axrep1d20130904r600} 
we have chosen $10^4$ points in the frequency interval $[0,100m_{\mathrm{e}}c^2]$, which means that $\Delta\omega_{\bm{K}}=0.01m_{\mathrm{e}}c^2$. The Fourier transform of 
the amplitudes (for the Thomson scattering see Eq.~\eqref{tt6}) is then calculated using the trapezoid algorithm for the integration. This means that we observe artificial revivals of 
$\tilde{\mathcal{A}}_{\mathrm{Th},\sigma}^{(+)}(\phi_{\mathrm{r}})$, separated by
\begin{equation}
\Delta\phi_{\mathrm{r}}=2\pi\frac{\omega_0}{\Delta\omega_{\bm{K}}}.
\label{tt12}
\end{equation}
Therefore, in order to make them well-separated from the real contribution we have to choose $\omega_0$ much larger than $\Delta\omega_{\bm{K}}$. 
In the numerical analysis presented above we have put $\omega_0=m_{\mathrm{e}}c^2$, which means that the closest artificial revivals appear for 
$\phi_{\mathrm{r}}\approx \pm 200\pi$, and by increasing $\omega_0$ we also increase their distance. This also puts the bound on $\phi_{\mathrm{max}}$; 
the maximum phase $\phi_{\mathrm{max}}$ should be large but cannot exceed $\Delta\phi_{\mathrm{r}}$. Lastly, it is important to note that two integrated 
distributions, Eqs.~\eqref{t28} and \eqref{tt11}, represent the same quantity and should give the same results.

In Fig.~\ref{puls3axcARd20130907} we present the energy distribution of radiation emitted in the direction $(\theta_{\bm{K}},\varphi_{\bm{K}})=(0.99999\pi,\pi)$, 
or equivalently for $(\Phi_{\bm{K}},\Theta_{\bm{K}})=(\pi/2,1.00001\pi)$. The distribution shows typical interference structures. The synthesis of 
this frequency pattern to the time-domain exhibits two main peaks together with two small sidelobes. The width of these peaks are of the order of 
$\delta\phi_{\mathrm{r}}\approx 0.1$ which means that they last for roughly $\delta t\approx 0.1/m_{\mathrm{e}}c^2=0.1 t_{\mathrm{C}}$, where $t_{\mathrm{C}}$ is the Compton time,
\begin{equation}
t_{\mathrm{C}}=\frac{\lambdabar_{\mathrm{C}}}{c}=\frac{1}{m_{\mathrm{e}}c^2}\approx 1.3\times 10^{-21}\mathrm{s}.
\label{comptontime}
\end{equation}
This time is many orders of magnitude smaller than the interaction-time of electrons with the laser pulse, which for the considered energy of electrons 
is a bit larger than $T_{\mathrm{p}}/2$.

\begin{figure}
\includegraphics[width=7.3cm]{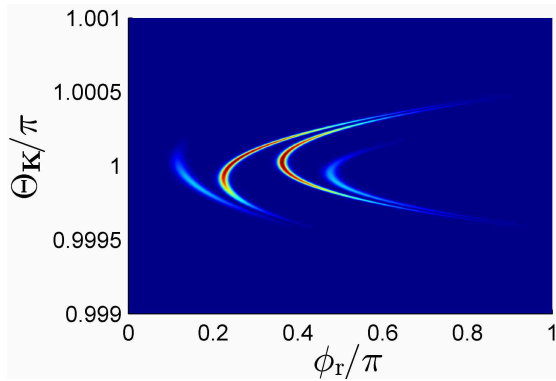}%
\caption{(Color online) Color map of the power distribution, Eq.~\eqref{tt9}, for the same geometry and parameters as in Fig.~\ref{surf3axrep1d20130904r600}, 
and for $\omega_0=m_{\mathrm{e}}c^2$.
\label{puls3axcAR2d20130907r600}}
\end{figure}
\begin{figure}
\includegraphics[width=7.3cm]{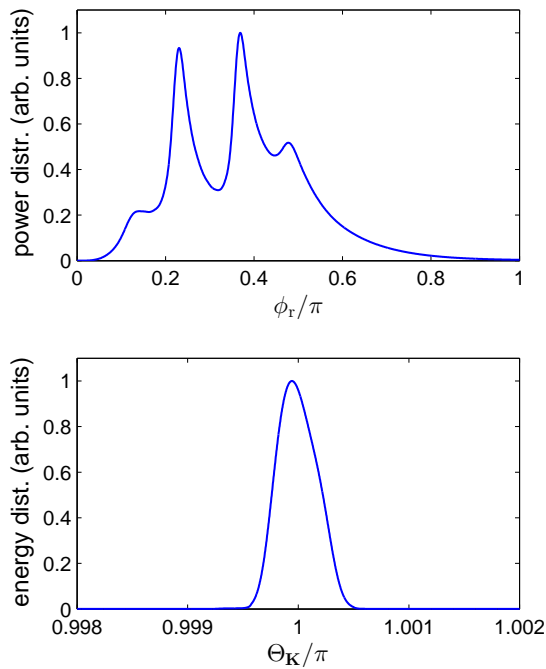}%
\caption{(Color online) Integrated over $\Theta_{\bm{K}}$ power distribution, Eq.~\eqref{tt10}, as a function of the retarded phase 
$\phi_{\mathrm{r}}$ for $\omega_0=m_{\mathrm{e}}c^2$ (upper panel). The lower panel shows the integrated energy distribution, Eqs.~\eqref{t28} and~\eqref{tt11}, 
as a function of $\Theta_{\bm{K}}$ for $\Phi_{\bm{K}}=\pi/2$. Both distributions are normalized to their maximum values, and the remaining parameters are the same as in Figs.~\ref{surf3axrep1d20130904r600} and \ref{puls3axcAR2d20130907r600}.
\label{puls3axcB2d20130907}}
\end{figure}

This clearly proves the coherent properties of the high-frequency supercontinuum generated in the laser-induced nonlinear Thomson (Compton) scattering. 
This supercontinuum can be synthesized to very short (zepto- or even yoctosecond) pulses. Let us also remark that within such a pulse the electric field does not 
oscillate, as it is presented in Fig.~\ref{puls3axcBd20130907}. This means that the emitted radiation is generated practically in the form of a 'one-cycle' pulse,
or even well-separated two 'half-cycle' pulses. This is of course the consequence of the three-cycle pulse used as a driving force.

It is well-known that the structure of the frequency distribution of emitted radiation is very sensitive to even a small change of emission angles 
(cf., Fig.~\ref{surf3axrep1d20130904r600}). Therefore, the question arises: How sensitive is the temporal power distribution of emitted radiation to such a change? 
To investigate this problem we shall consider, as above, the emission in the scattering plane. The synthesis of the energy spectrum shown in Fig.~\ref{surf3axrep1d20130904r600} 
leads to the temporal power distribution presented as the color map in Fig.~\ref{puls3axcAR2d20130907r600}. We observe that radiation is emitted in a form of sharp 
stripes and that, even after integrating with respect to $\Theta_{\bm{K}}$, the main temporal peaks from Fig.~\ref{puls3axcARd20130907} show up for the same $\phi_{\mathrm{r}}$.
Note that this happens with a smaller contrast, as presented in the upper frame of Fig.~\ref{puls3axcB2d20130907}. In the lower frame of Fig.~\ref{puls3axcB2d20130907} we also depict 
the partially integrated energy distribution, Eqs.~\eqref{t28} and \eqref{tt11}. It shows that, as expected, the high-frequency radiation is emitted in a very narrow cone. 
This explains why the very sharp temporal structure survives the integration over the emission angles.

\section{Conclusions}
\label{conclusions}

An appearance of a broad bandwidth radiation (spanning a few MeV), which is sharply elongated around the propagation
direction of the electron beam, has been demonstrated from nonlinear Thomson (Compton) scattering.
Our analysis of temporal distributions of the observed radiation shows that it can be used for a synthesis
of zeptosecond (likely even yoctosecond) pulses. Note that this is possible provided that the broad bandwidth radiation is coherent,
which clearly proves that nonlinear Thomson or Compton scattering can lead to a generation of a supercontinuum.

When analyzing properties of the formed zeptosecond pulses, we discovered that these are one-cycle (half-cycle) pulses. 
We have also seen that Thomson (Compton) radiation is very sensitive to a change of 
its emission direction. However, as we showed in this paper, the ultra-short pulses survive the space averaging. 
In light of this fact, we conclude that the Thomson (Compton) process can be used as a novel source of
zeptosecond (yoctosecond) pulsed radiation. This, in further perspective, will enable the entrance to new physical
regimes of intense laser physics.

For convenience, we based our analysis of ultra-short pulse generation 
presented in this paper on a classical theory of nonlinear Thomson scattering. However, as we argued in Ref.~\cite{KKscale}, 
for an appropriate choice of parameters, the Thomson and Compton spectra coincide 
in an entire frequency range. This is exactly the case considered in this paper. While in Ref.~\cite{KKscale}
we investigated a scaling law for Thomson and Compton spectra and discussed it, for the first time, in the context of 
quantum recoil of scattered electrons, as well as the spin and polarization effects for arbitrary short laser pulses, 
here we demonstrated and explained other features of emitted radiation such as
oscillatory and pulsating patterns exhibited by both frequency spectra.

We note that the Thomson process can be only considered as an approximation of the fundamental
Compton scattering, which describes the actual physical situation with the electron spin degrees of freedom included.
Therefore, it is important to analyze not only similarities but also differences between these two approaches.
These problems have been studied in Ref.~\cite{KKscale} and they will also be discussed elsewhere in the context of pulse generation.

\section*{Acknowledgments}
This work is supported by the Polish National Science Center (NCN) under Grant No. 2012/05/B/ST2/02547.


\begin{thebibliography}{99}

\bibitem{Tajima}
T. Tajima and J. M. Dawson, Phys. Rev. Lett. {\bf 43}, 267 (1979).

\bibitem{Phuoc}
K. Ta Phuoc, S. Corde, C. Thaury, V. Malka, A. Tafzi, J. P. Goddet, R. C. Shah, S. Sebban, and A. Rousse, Nat. Photonics {\bf 6}, 308 (2012).

\bibitem{Umstadter}
D. Umstadter, J. Phys. D: Appl. Phys. {\bf 36}, R151 (2003).

\bibitem{review1} 
F. Ehlotzky, K. Krajewska, and J. Z. Kami\'nski, Rep. Prog. Phys. \textbf{72}, 046401 (2009).

\bibitem{review2} 
A. Di Piazza, C. M\"uller, K. Z. Hatsagortsyan, and C. H. Keitel, Rev. Mod. Phys. {\bf 84}, 1177 (2012).

\bibitem{ELI}
http://www.extreme-light-infrastructure.eu

\bibitem{Neville} 
R. A. Neville and F. Rohrlich, Phys. Rev. D {\bf 3}, 1692 (1971).

\bibitem{Narozhny1}
N. B. Narozhny and M. S. Fofanov, JETP {\bf 83}, 14 (1996).

\bibitem{Boca2009}
M. Boca and V. Florescu, Phys. Rev. A {\bf 80}, 053403 (2009).

\bibitem{Mackenroth}
F. Mackenroth, A. Di Piazza, and C. H. Keitel, Phys, Rev. Lett. {\bf 105}, 063903 (2010).

\bibitem{Mack}
F. Mackenroth and A. Di Piazza, Phys. Rev. A {\bf 83}, 032106 (2011).

\bibitem{Seipt2011}
D. Seipt and B. K\"ampfer, Phys. Rev. A {\bf 83}, 022101 (2011).

\bibitem{Boca2011}
M. Boca and V. Florescu, Eur. Phys. J. D {\bf 61}, 449–462 (2011).

\bibitem{Boca2012}
M. Boca, V. Dinu, and V. Florescu, Phys. Rev. A {\bf 86}, 013414 (2012).

\bibitem{KKcompton}
K. Krajewska and J. Z. Kami\'nski, Phys. Rev. A {\bf 85}, 062102 (2012).

\bibitem{Macken}
F. Mackenroth and A. Di Piazza, Phys. Rev. Lett. {\bf 110}, 070402 (2013).

\bibitem{KKpol}
K. Krajewska and J. Z. Kami\'nski, Laser Part. Beams {\bf 31}, 503 (2013).

\bibitem{KKscale}
K. Krajewska and J. Z. Kami\'nski, arXiv:1308.1663.

%\bibitem{Narozhny2}
%N. B. Narozhny and M. S. Fofanov, Laser Phys. {\bf 7}, 141 (1997).

%\bibitem{Heinzl}
%T. Heinzl, A. Ilderton, and M. Marklund, Phys. Lett. B {\bf 692}, 250 (2010).

%\bibitem{Titov}
%A. I. Titov, H. Takabe, B. K\"ampfer, and A. Hosaka, Phys. Rev. Lett. {\bf 108}, 240406 (2012).

\bibitem{KMK2013}
K. Krajewska, C. M\"uller, and J. Z. Kami\'nski, Phys. Rev. A {\bf 87}, 062107 (2013).

\bibitem{Rash1}
S. P. Roshchupkin, A. A. Lebed', E. A. Padusenko, and A. I. Voroshilo, Laser Phys. {\bf 22}, 1113 (2012).

\bibitem{Rash2}
S. P. Roshchupkin, A. A. Lebed', and E. A. Padusenko, Laser Phys. {\bf 22}, 1513 (2012).

\bibitem{Boca2013}
M. Boca, Cent. Eur. J. Phys., DOI: 10.2478/s11534-013-0287-0 (2013).

\bibitem{Rash3}
A. A. Lebed' and S. P. Roshchupkin, Laser Phys. {\bf 23}, 125301 (2013).

\bibitem{Ghebregziabher}
I. Ghebregziabher, B. A. Shadwick, and D. Umstadter, Phys. Rev. ST Accel. Beams {\bf 16}, 030705 (2013).

\bibitem{Zhang}
K. Zhao, Q. Zhang, M. Chini, Y. Wu, X. Wang, and Z. Chang, Optics Lett. {\bf 37}, 3891 (2012). 

\bibitem{HHG11}
M. Ferray, A. L'Huillier, X. F. Li, L. A. Lompr\'e, G. Mainfray, and C. Manus, J. Phys. B {\bf 21}, L31 (1988).

\bibitem{HHG22}
A. McPherson, G. Gibson, H. Jara, U. Johann, T. S. Luk, I. A. McIntyre, {\it et al.}, J. Opt. Soc. Am. B {\bf 4}, 595 (1987).

\bibitem{Farkas}
Gy. Farkas and Cs. T\'oth, Phys. Lett. A {\bf 168}, 447 (1992).

\bibitem{Keitel}
M. C. K\"ohler, T. Pfeifer, K. Z. Hatsagortsyan, and C. H. Keitel, Adv. At. Mol. Opt. Phys. {\bf 61}, 159 (2012).

\bibitem{Shan}
B. Shan and Z. Chang, Phys. Rev. A {\bf 65}, 011804(R) (2001).

\bibitem{Jaron}
C. Hern\'andez-Garc\'ia, J. A. P\'erez-Hern\'andez, T. Popmintchev, M. M. Murnane, H. C. Kapteyn,
A. Jaro\'n-Becker, A. Becker, and L. Plaja, Phys. Rev. Lett. {\bf 111}, 033002 (2013).

\bibitem{FEL}
B. W. J. McNeil and N. R. Thompson, Nature Photon. {\bf 4}, 814 (2010). 

\bibitem{Dunning}
D. J. Dunning, B. W. J. McNeil, and N. R. Thompson, Phys. Rev. Lett. {\bf 110}, 104801 (2013).

\bibitem{Jackson1975}
J. D. Jackson, \textit{Classical Electrodynamics} (John Wiley and Sons, New York, 1975).

\bibitem{LL2}
L. D. Landau and E. M. Lifshitz, \textit{The Classical Theory of Field} (Butterworth-Heinemann, Oxford, 1987).

\bibitem{Salamin}
Y. I. Salamin and F. H. M. Faisal, Phys. Rev. A {\bf 54}, 4383 (1996).

\bibitem{Hartem}
F. V. Hartemann and A. K. Kerman, Phys. Rev. Lett. {\bf 76}, 624 (1996).

\bibitem{Hart}
F. V. Hartemann, {\it High-Field Electrodynamics} (CRC Press, Boca Raton, FL, 2002).

\bibitem{KKasymm}
K. Krajewska and J. Z. Kami\'nski, Phys. Rev. A {\bf 86}, 021402(R) (2012).

\bibitem{KKcomb}
K. Krajewska and J. Z. Kami\'nski, arXiv:1307.5433.

\bibitem{super1}
R. R. Alfano and S. L. Shapiro, Phys. Rev. Lett. {\bf 24}, 584 (1970); \textit{ibid.}, 592 (1970).

\bibitem{super2}
J. M. Dudley, G. Genty, and S. Coen, Rev. Mod. Phys. {\bf 78}, 1135 (2006).

\bibitem{super3}
B. R. Washburn, S. A. Diddams, N. R. Newbury, J. W. Nicholson, M. F. Yan, and C. G. J{\o}rgensen, Opt. Lett. {\bf 29}, 250 (2004).

\bibitem{super4}
W. Drexler, U. Morgner, F. X. K\"artner, C. Pitris, S. A. Boppart, X. D. Li, E. P. Ippen, and J. G. Fujimoto, Opt. Lett. {\bf 24}, 1221 (1999).

\bibitem{super5}
R. T. Neal, M. D. C. Charlton, G. J. Parker, C. E. Finlayson, M. C. Netti, and J. J. Baumberg, Appl. Phys. Lett. {\bf 83}, 4598 (2003).

\bibitem{super6}
G. Brambilla, F. Koizumi, V. Finazzi, and D. J. Richardson, Electron. Lett. {\bf 41}, 795 (2005).

\bibitem{KKbreit}
K. Krajewska and J. Z. Kami\'nski, Phys. Rev. A {\bf 86}, 052104 (2012).

\bibitem{HHG}
F. Krausz and M. Ivanov, Rev. Mod. Phys. {\bf 81}, 163 (2009).

\bibitem{HHG1}
R. A. Ganeev, Laser Phys. Lett. {\bf 9}, 175 (2012).

\bibitem{HHG2}
F. H. M. Faisal and J. Z. Kami\'nski, Phys. Rev. A {\bf 54}, R1769 (1996).

\end{thebibliography}
\end{document}